# Some problems of the projectile motion with a square-law resistance

## Peter Chudinov

*Department of Engineering, Perm State Agro-Technological University, 614990, Perm, Russia*

*chupet@mail.ru*

The influence of the force of the quadratic resistance of the medium on the change in some interesting characteristics of the motion of the projectile, which take place when the projectile moves in vacuum, is investigated. Loci that ensure maximization of the flight range, the arc length of the projectile trajectory and a non-decreasing of the length of the radius-vector are constructed numerically (and partly analytically). As examples, the motion of a baseball, a tennis ball and a badminton shuttlecock is studied.

## 1. Introduction

The study of the motion of a projectile thrown at an angle to the horizon is a classic task of science and is included in introductory physics courses in colleges and universities. In the absence of resistance of the medium, this problem in all its aspects has been thoroughly investigated and is presented in various textbooks. Even schoolchildren know that to achieve the longest range one must to throw the ball at a 45° angle. However, not everyone knows that besides the aforementioned wonderful angle, there are other interesting throwing angles in this problem. We present them in the Table 1 in ascending order of the angle value. In this case, we will keep in mind that the indicated values were obtained when the projectile was moving in a vacuum.

**Table 1.** Remarkable angles of throw when the projectile moves in a vacuum.

| №№ | remarkable throwing angle | The property that provides a given angle of throw |
|---|---|---|
| 1 | $\theta_0 = 45°$ | Maximum range at a given speed |
| 2 | $\theta_0 = 56.46°$ | Maximum arc length of the projectile trajectory |
| 3 | $\theta_0 = 70.53°$ | Separation of the nature of the change in the length of the radius-vector |
| 4 | $\theta_0 = 90°$ | Maximum height at a given speed |

For brevity, we will call these angles 1, 2, 3, 4 angles. For example, hereinafter, the term "angle 1" means the initial throwing angle of the projectile, that provides the maximum range of its motion. The term "angle 2" means the initial throwing angle of the projectile, that provides the maximum length of the arc of its trajectory. The term "angle 3" means the initial throwing angle, which separations the nature of the change in the length of the radius-vector of the projectile. Angle 4 is not covered in this article. The purpose of this work is to study the effect of the quadratic resistance of the medium on the change in angles 1, 2 and 3.

We explain the appearance and meaning of angle 2. Let $V_0$ − initial velocity of the projectile, $g$ − acceleration of gravity. It is known [1,5], that when the projectile moves in vacuum, the arc length of the trajectory $S_{tr}$ as a function of the throwing angle $\theta_0$ is determined by the formula

$$S_{tr}(\theta_0) = \frac{V_0^2}{g}\left(\sin\theta_0 + \cos^2\theta_0 \ln\left(\frac{1+\sin\theta_0}{\cos\theta_0}\right)\right). \tag{1}$$

For a fixed initial velocity $V_0$, this function has a local maximum at $\theta_0 = 56.46°$. This angle provides the maximum arc length of the projectile trajectory.

Now let's describe the angle 3 when the projectile moves in vacuum. In [2] Walker described an interesting property of motion - the nature of the change in the length of the radius-vector depending on the magnitude of the throwing angle. Fig. 1 from paper [2] illustrates this property. The following notation is used here: $r(x) = \sqrt{x^2 + y(x)^2}$ is the length of the radius-vector of the projectile, $x, y$ – are the Cartesian coordinates of the projectile. In the figure, for the initial speed $V_0 = 40$ m/s, a family of curves $r = r(x)$ is constructed at various angles of throw in the range from $\theta_0 = 62.5°$ to $\theta_0 = 80°$. Each curve describes the movement of the projectile from the point of launch to the point of impact, where $r = x$. It can be seen from the figure that the nature of the change in the parameter $r$ depends on the value of the throwing angle. For example, at an angle of throw $\theta_0 = 62.5°$, the parameter $r$ increases monotonically throughout the entire flight of the projectile. On the contrary, at an angle of throw $\theta_0 = 80°$, the value of $r$ at first increases, then decreases, and then increases again before the point of incidence. There is a certain critical angle of throw $\theta_0^{cr} = 70.53°$ that separates these types of behavior of the parameter $r$ and does not depend on the initial velocity $V_0$ (angle 3). This critical angle is determined from the condition of the existence of a *single* root of the equation

$$\frac{dr}{dt} = 0. \tag{2}$$

Fig. 1 curve for the angle $\theta_0^{cr}$ is shown in red. Plots of functions $r = r(x), y = y(x)$ are plotted in Fig. 2. for the throw angle $\theta_0^{cr}$. From Fig. 2 we can see that angle 3 always corresponds to the position of the radius-vector on the descending branch of the projectile trajectory.

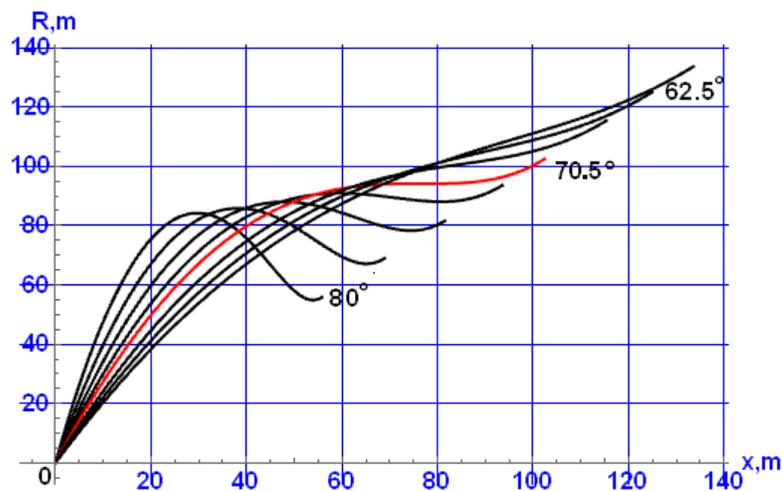

Figure 1. Graphs of $r = r(x)$ function.

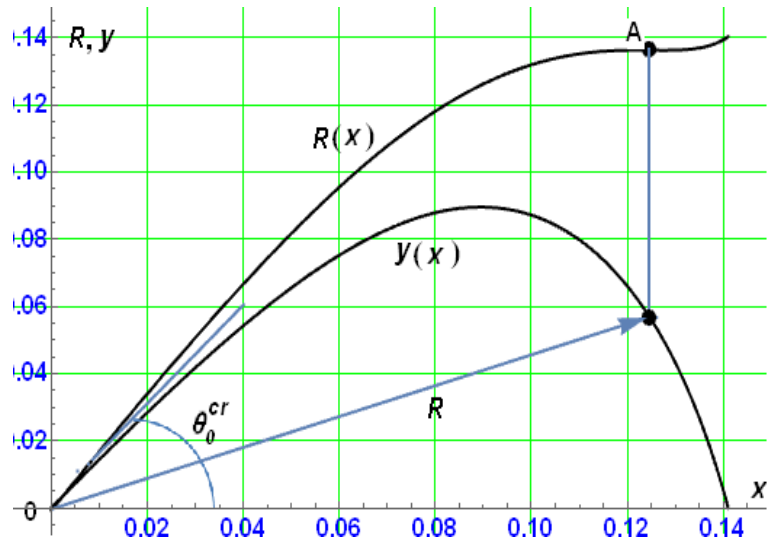

Figure 2. Graphs of $r = r(x), y = y(x)$ functions.

Recently, studies have appeared [6,7], in which the influence of the forces of resistance of the medium on the change in the characteristics of motion given in Table 1 is considered. The article [6] numerically studies the effect on the change in angle 2 of resistance forces, which have a linear and quadratic dependence on speed. For the case of a quadratic dependence of the force of resistance of the medium on the velocity of the projectile, a general conclusion is made about the increase in angle 2 in comparison with the case of the absence of resistance. For the case of linear resistance, the dependence $S_{tr} = S_{tr}(\theta_0)$ is approximated by a cubic polynomial relative to the projectile throwing angle. In [7], the influence of only the linear resistance of the medium on the change of angle 3 is considered. The results indicate a decrease of angle 3 with an increase of the coefficient of resistance of the medium. This article follows the general direction of research in [6,7], but only the case of the quadratic resistance of the medium is considered. The purpose of this research is to study in more detail the effect of the quadratic resistance on the change in angles 1, 2 and 3.

2. **Study of the effect of square-law resistance on the change in angle 1**

Let the projectile move in a medium with a square law of resistance (Fig. 3).

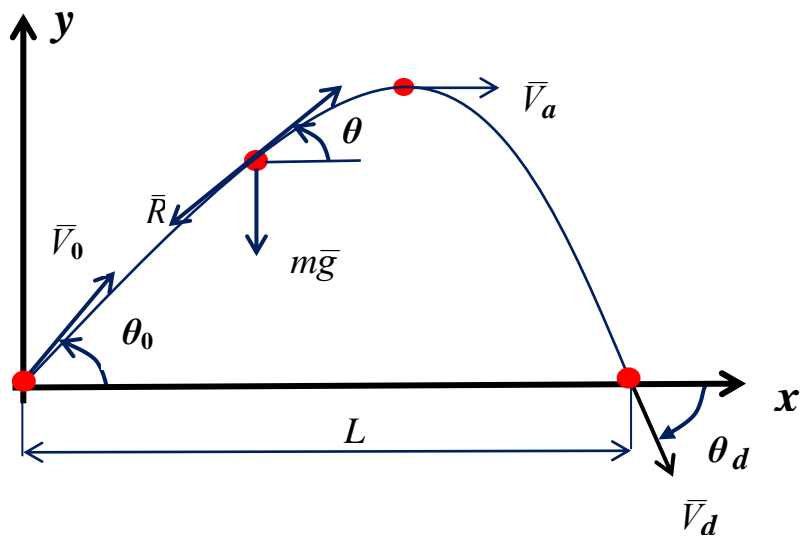

Figure 3. Some characteristics of the projectile motion.

Let's write down the equations of the projectile motion. Let us consider the motion of a projectile with mass $m$ launched at an angle $\theta_0$ with an initial speed $V_0$ under the influence the force of gravity and resistance force $R = mgkV^2$. Here $g$ is the acceleration of gravity, $k$ is the drag constant and $V$ is the speed of the object. Air resistance force $R$ is proportional to the square of the speed of the projectile and is directed opposite the velocity vector. It is assumed that the projectile is at the origin at the initial instant and the point of impact of the projectile lies on the same horizontal $y = 0$ (see Fig. 3). In ballistics, the movement of a projectile is often studied in projections on natural axes [3]. The equations of the projectile motion in this case have the form

$$\frac{dV}{dt} = -g\sin\theta - gkV^2, \quad \frac{d\theta}{dt} = -\frac{g\cos\theta}{V}, \quad \frac{dx}{dt} = V\cos\theta, \quad \frac{dy}{dt} = V\sin\theta. \tag{3}$$

Here $\theta$ is the angle between the tangent to the trajectory of the projectile and the horizontal, $x, y$ are the Cartesian coordinates of the projectile. The drag coefficient $k$ is usually determined through the terminal velocity of the projectile: $k = 1/V_{term}^2$.

When the projectile moves in a vacuum, the value of angle 1 does not depend on the initial velocity $V_0$; when moving in a medium with resistance, angle 1 depends on both $V_0$ and the value of the drag coefficient $k$. The parameters $V_0, k$ form the dimensionless parameter $p = kV_0^2$. It has a clear physical meaning - it is the ratio of the resistance force of the medium to the weight of the projectile at the moment the movement starts. Therefore, when studying the effect of quadratic resistance on the value of angle 1, we will investigate the dependence of angle 1 on the parameter $p$. This can be done in two ways - numerical and analytical. The first method is the numerical integration of the differential equations of projectile motion (3). The second way is the use of high-precision analytical approximations [8,9] describing the movement of the projectile. In this research, both methods are used. In [9], an explicit formula was obtained for the projectile flight range $L$ (see Fig. 3) as a function of three parameters: $L = L(\theta_0, V_0, k)$. The maximum of this function $L_{max}$ at a given value of the parameter $p = kV_0^2$ determines the sought angle 1. The final result is shown in Fig. 4. This figure on the plane of parameters $(p, \theta_0)$ shows the locus of points for which the condition $L = L_{max}$ is satisfied. The solid red line in Fig. 4 was obtained using formulas [9], the dotted curve was obtained by numerical integration of the system of differential equations (3) by the Runge-Kutta method of the 4th order. We note the high accuracy of the formula obtained in [9] for the range.

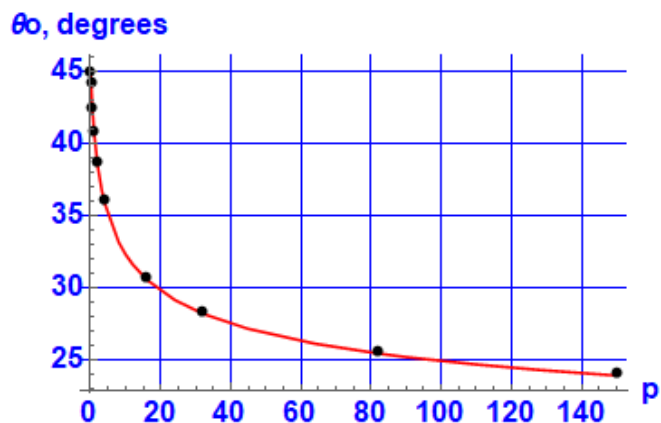

Figure 4. The locus of the points, for which $L = L_{max}$, on the plane $(p, \theta_0)$.

The curve in Fig. 4 allows to immediately determine the desired angle 1 at the given values of $V_0, k$ without calculations.

## 3. Study of the effect of square-law resistance on the change in angle 2

Let us study the influence of the resistance of the medium on the change in the angle 2. In the case of quadratic resistance, the arc length of the trajectory is determined by the following ultimate formula [4]:

$$S_{tr} = \frac{1}{2gk}\ln\left(1+kV_0^2\cos^2\theta_0\left(f(\theta_0)-f(\theta_d)\right)\right), \quad \text{where} \quad f(\theta) = \frac{\sin\theta}{\cos^2\theta} + \ln\left(\tan\left(\frac{\theta}{2}+\frac{\pi}{4}\right)\right). \quad (4)$$

In the formula (4) $\theta_d$ is the angle of incidence of the projectile (see Fig. 3). In contrast to formula (1), in which equality takes place $\theta_d = -\theta_0$, in this case the value of the angle of incidence is not known in advance and must be calculated using the methods mentioned. The final result is shown in Fig. 5, 6. Figure 5 shows graphs of the dependence of the normalized arc length $S$ on the initial throwing angle $\theta_0$ for various values of the resistance coefficient $k$. The arc length $S$ is made dimensionless by introducing a normalization factor $n$:

$$S(\theta_0) = S_{tr}\cdot n, \quad \text{where} \quad n = \frac{g}{V_0^2}.$$

Figure 5 contains five curves. Red curve 1 was obtained at $k = 0$, curve 2 was obtained at $k = 0.000625$ $s^2/m^2$, curve 3, at $k = 0.002$ $s^2/m^2$, and curve 4, at $k = 0.022$ $s^2/m^2$. The value of the initial speed is the same for curves 1, 2, 3, 4: $V_0 = 40$ m/s. Curve 1 corresponds to the movement of the projectile in vacuum, curves 2, 3, 4 at the used values of the drag coefficient $k$ describe the motion of a baseball, a tennis ball and a badminton shuttlecock in the air, respectively. Solid black lines on curves 2,3,4 were obtained by numerical integration of the system of equations of projectile motion (3) by the Runge-Kutta method of the 4th order. Dotted red lines on the same curves were obtained by using analytical approximations [8]. The complete coincidence of the solid and dotted curves testifies to the high accuracy of the approximations [8]. From the analysis of curves 1, 2, 3, 4 it can be seen that with an increase in the resistance coefficient $k$, angle 2 enlarges. Blue curve 5 is obtained numerically and represents the locus of points for which the arc length of the corresponding trajectory is maximum, on the plane $(S, \theta_0)$. For a more detailed analysis, we will construct the locus of points corresponding to one condition or another, on different planes of parameters. This remark applies to both angle 2 and angle 3. In other words, for any given values of the initial velocity of the projectile $V_0$ and the drag coefficient $k$, the values of the parameter $S_{max}$ and angle 2 lie on curve 5. The start and end points of curve 5 have coordinates (56.46°, 1.2) and (90°, 0). Directly from Fig. 5 we can see that curve 5 passes through the points of maximum of the curves $S = S(\theta_0)$. This means that for all points of curve 5 the condition is satisfied

$$\frac{dS(\theta_0)}{d\theta_0} = 0.$$

We use the parameter $p = kV_0^2$ again. In Fig. 6 the same curve 5 is shown on the parameter plane $(p, \theta_0)$. Points 1 and 2 on the curve correspond to the movement of the tennis ball and the badminton shuttlecock, respectively. Motion parameters for tennis are $V_0 = 70$ m/s, $k = 0.002$ $s^2/m^2$, for shuttlecock – $V_0 = 80$ m/s, $k = 0.022$ $s^2/m^2$. The curve in Fig. 6 allows to immediately determine the desired angle 2 at the given values of $V_0, k$ without calculations. The start and end points of the curve have coordinates $(0, 56.46°)$ and $(160, 81.43°)$.

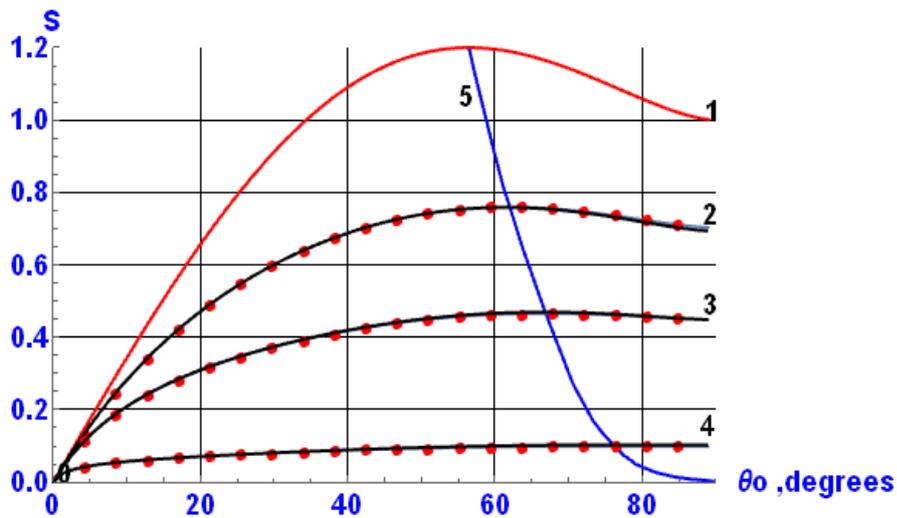

Figure 5. Graphs of $S = S(\theta_0)$ functions.

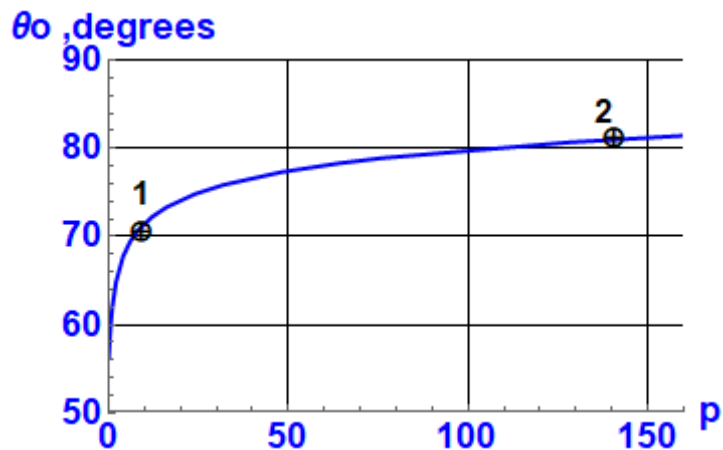

Figure 6. The locus of the points, for which $S = S_{max}$, on the plane $(p, \theta_0)$.

### 4. Investigation of the effect of square-law resistance on the change in angle 3

Let's turn to the study of angle 3. To calculate the length of the radius-vector, you need to know the Cartesian coordinates $x, y$ of the projectile at each point of its trajectory. We will calculate them in the two ways mentioned above. Let us investigate how the critical throwing angle changes when a projectile with a square-law resistance moves. First, we consider the effect of the drag coefficient $k$ on angle 3. In Fig. 7, a family of curves $\theta_0^{cr} = \theta_0^{cr}(k)$ is numerically constructed for various values of the initial velocity.

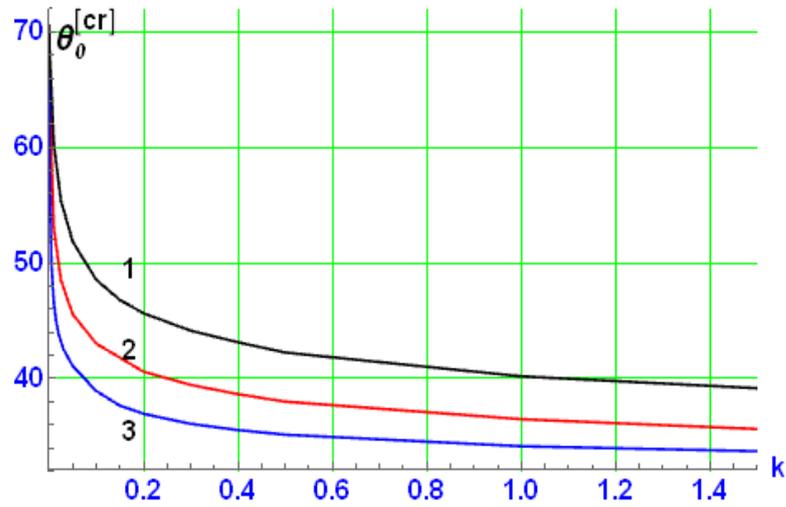

Figure 7. Dependence $\theta_0^{cr} = \theta_0^{cr}(k)$.

Curve 1 is plotted at $V_0 = 20$ m/s, curve 2 - at $V_0 = 40$ m/s, curve 3 - at $V_0 = 80$ m/s. From the figure it follows that with an increase in the value of the resistance coefficient, the value of angle 3 decreases. For the convenience of the study, we make the quantities $r$ and $x$ dimensionless: $R = r \cdot n$, $x = x \cdot n$. In Fig. 8, a family 1,2,3 curves $R = R(x)$ are constructed at the same values of the initial velocities as in Fig. 5 and for critical values of the angle of throw $\theta_0^{cr}$.

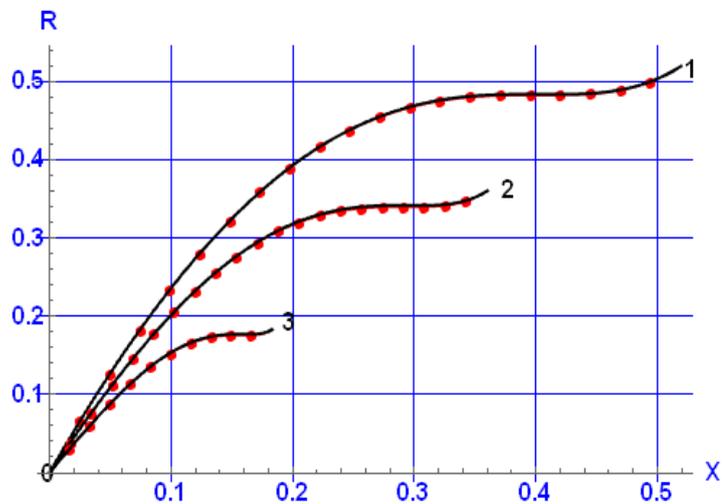

Figure 8. Dependence $R = R(x)$.

Parameter $k$ is constant: $k = 0.001$ s$^2$/m$^2$. It can be seen from the figure that with an increase in the initial speed, angle 3 decreases. For curve 1 it is equal to: $\theta_0^{cr} = 68.5°$, for curve 2: $\theta_0^{cr} = 64.5°$, for curve 3: $\theta_0^{cr} = 57.9°$. Solid black lines on curves 1, 2, 3 were obtained by numerical integration of the system of equations of projectile motion (2) by the Runge-Kutta method of the 4th order. Dotted red lines on the same curves were obtained by using analytical approximations [8].

The final result is shown in Fig. 9. The locus of points for which condition (2) is satisfied is numerically constructed on the parameter plain $\left(\theta_0^{cr}, p\right)$. In fact, this is the locus of the points of inflection of the length $r(x)$ of the radius-vector of the projectile (point A in Fig. 2). For any given values of the initial velocity of the projectile $V_0$ and the drag coefficient $k$, the value of angle 3 will lie on the resulting curve. The start and end points of the curve have coordinates (0, 70.53°) and (160, 43.2°). Points numbered 1, 2, 3 correspond to the movement of a baseball, a tennis ball and a badminton shuttle, respectively.

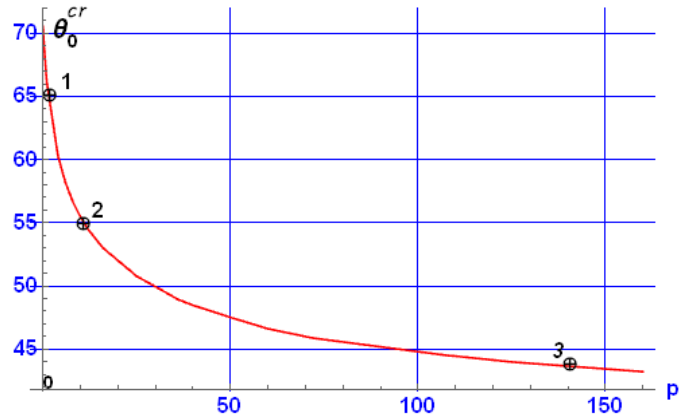

Figure 9. Locus of points meeting a condition $\dfrac{dr}{dt} = 0,$ on a plane $\left(p, \theta_0^{cr}\right)$.

The characteristics of the movement of these sports equipment are shown in Table 2.

**Table 2. Characteristics of the motion of sports equipment.**

| sports equipment | $V_0$, m/s | $k$, s$^2$/m$^2$ | $p = kV_0^2$ | $\theta_0^{cr}$ |
|---|---|---|---|---|
| baseball | 50 | 0.000625 | 1.56 | 64.6° |
| tennis | 70 | 0.002 | 9.8 | 55.6° |
| badminton | 80 | 0.022 | 140.8 | 43.5° |

Let us now prove an interesting property of angles 3. Angles 3 are found from the condition $dr/dt = 0$. Let us differentiate the radius-vector with respect to time taking into account equations (3):

$$\frac{dr}{dt} = \frac{x\dfrac{dx}{dt} + y\dfrac{dy}{dt}}{r} = \frac{xV\cos\theta + yV\sin\theta}{r} = \frac{V}{r}(x\cos\theta + y\sin\theta) = 0.$$

This implies the equality $x\cos\theta + y\sin\theta = 0$, or $\tan\theta = -x/y$. From the system of equations of projectile motion (3), we obtain the relation $\tan\theta = dy/dx.$ Comparing these two relations, we obtain the differential equation

$$\frac{dy}{dx} = -\frac{x}{y},$$

whose solution has the form

$$(x - x_0)^2 + (y - y_0)^2 = a^2. \tag{5}$$

This means that the Cartesian coordinates $x, y$ of the inflection points of the radius-vector length $r(x)$, obtained from the condition $dr/dt = 0$, for any values of the drag coefficient $k$ lie on a circular arc of the form (5). Point $O_1$ (the center of the circle) has coordinates $O_1(x_0, y_0)$, parameter $a$ is the radius of the circle.

Let $\bar{x}, \bar{y}$ be dimensionless coordinates: $\bar{x} = nx$, $\bar{y} = ny$. Let's keep the previous designations for them: $x, y$. It is difficult to determine the parameters $x_0, y_0, a$ analytically. Therefore, we will do the opposite - we will construct sets of points $(x, y)$ and for them we will select the corresponding equation of the circle. In Fig. 10, the geometric location of the points corresponding to condition (2) is plotted on the plane $(x, y)$. Red points were obtained numerically, solid black line represent an arc of a circle of the form (5) with dimensionless parameters:

$$x_0 = -0.489474, \quad y_0 = 1.19222, \quad a = 1.28879.$$

The resistance coefficient $k$ on curve runs through all values of the interval $[0, \infty)$. The upper end of arc correspond to movement without resistance, the lower to movement with very high resistance. Fig. 10 is universal. It contains $x, y$ coordinates, corresponding to the angles 3, for any values of speed $V_0$ and drag coefficient $k$.

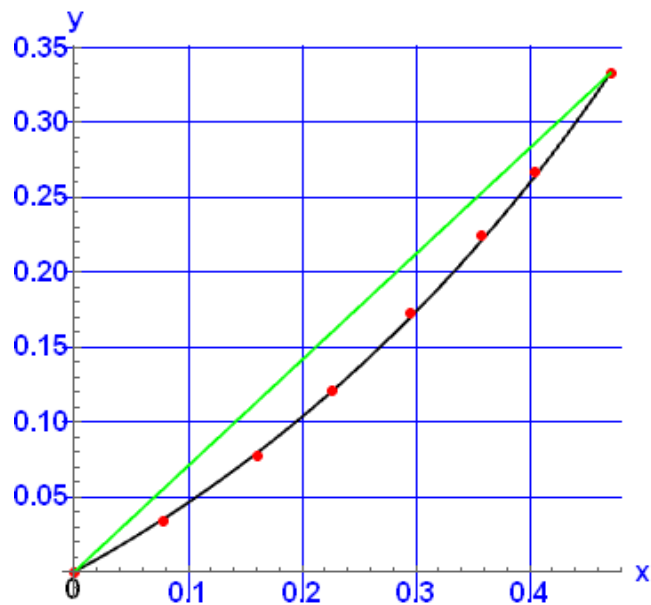

Figure 10. Locus of points meeting a condition $\dfrac{dr}{dt} = 0,$ on a plane $Oxy.$

The upper and lower ends of the arc also lie on the straight line (shown in green). This straight line is given be the equation

$$y = \frac{\sqrt{2}}{2} x. \tag{6}$$

The angle of inclination of the line (6) to the $x$ axis is $35.26°$. Equation (6) can be deduced from the formulas of the article [2].

## 5. Conclusion

Thus, in this research we investigated the influence of the force of the quadratic resistance of the medium on the change in some interesting characteristics of the motion of the projectile, which take place when the projectile moves in vacuum. Numerically (and partly analytically) geometrical places of points are constructed that ensure maximization of the range, arc length of the projectile trajectory and a non-decreasing of the length of the radius-vector. Let us emphasize the diverse influence of the force of quadratic resistance on the magnitude of the studied angles 1, 2 and 3. With an increase of the value of the resistance force, the value of angle 2 increases, and the values of angles 1 and 3 decreases. It should also be noted the usefulness of the formulas [8,9], which made it possible to significantly reduce the amount of numerical calculations in this research.

## References

1. H. Sarafian, On projectile motion, *The Physics Teacher*, **37**, 86 (1999), https://dx.doi.org/10.1119/1.880184
2. J. Walker, Projectiles: are they coming or going?, *The Physics Teacher*, **33**, 282 (1995), http:// dx.doi.org/10.1119/1.2344221
3. B. Okunev, *Ballistics*, Vol.2, (Voyenizdat, Moscow, 1943), p.14
4. A. Tan, C. Frick and O. Castillo, The fly ball trajectory: an older approach revisited, *American Journal of Physics*, **55**, (1987), 37-40, https://doi.org/10.1119/1.14968
5. M. Karadag, A study for determining the launch angle that maximises the total distance travelled by the projectile during its flight in the projectile motion, *Physics Education*, **55,** (2020), 033011, https://doi.org/10.1088/1361-6552/ab7f97
6. H. Sarafian, What projective angle makes the arc-length of the trajectory in a resistive media maximum? A reverse engineering approach, *American Journal of Computational Mathematics*, **11,** (2021), 71-82, https://doi.org/ 10.4236/ajcm.2021.112007
7. W. Ribeiro and J. de Sousa, Projectile motion: the "coming and going" phenomenon, *The Physics Teacher*, **59**, 168 (2021), https://doi.org/10.1119/10.0003656
8. P. Chudinov, V. Eltyshev, Y. Barykin, Analytical construction of the projectile motion trajectory in midair, *MOMENTO Revista de Fisica,* **62,** (2021), 79-96, https://doi.org/10.15446/mo.n62.90752
9. P. Chudinov, Study of the projectile motion in midair using simple analytical formulas, *arXiv preprint arXiv:2103.11111* (2021)